\documentclass[prl,aps,showpacs,showkeys,superscriptaddress,twocolumn,dvips]{revtex4}

\usepackage{graphicx,color,amsmath,amssymb} 
\usepackage{hyperref}% PDF links

\def\vecDveceenp{$^{2}\vec{\rm H}(\vec{\rm e},{\rm e}'{\rm n}){\rm p}$ $ $}
\def\vecDveceepn{$^{2}\vec{\rm H}(\vec{\rm e},{\rm e}'{\rm p}){\rm n}$ $ $}

\def\vecDveceep{$^{2}\vec{\rm H}(\vec{\rm e},{\rm e}'{\rm p})$ $ $}

\def\Deenp{$^{2}{\rm H}({\rm e},{\rm e}'{\rm n}){\rm p}$ $ $}
\def\Deen{$^{2}{\rm H}({\rm e},{\rm e}'{\rm n})$ $ $}

\def\eep{$({\rm e},{\rm e}'{\rm p})$ $ $}
\def\een{$({\rm e},{\rm e}'{\rm n})$ $ $}

\begin{document}

\title{\boldmath The Charge Form Factor of the Neutron at Low Momentum
  Transfer from the $^{2}\vec{\rm H}(\vec{\rm e},{\rm e}'{\rm n}){\rm p}$
  Reaction} 

\author{E.~Geis}
\thanks{Reported results are based on the Ph.D. theses of E.G. and V.Z.}
\affiliation{Arizona State University, Tempe, AZ 85287}
\author{M.~Kohl}
\altaffiliation[Corresponding author, 
email ]{kohlm@jlab.org}
%\email{kohlm@jlab.org}
\affiliation{Laboratory for Nuclear Science and Bates Linear Accelerator
  Center, Massachusetts Institute of Technology, Cambridge, MA 02139} 
\author{V.~Ziskin}
\thanks{Reported results are based on the Ph.D. theses of E.G. and V.Z.}
\affiliation{Laboratory for Nuclear Science and Bates Linear Accelerator
  Center, Massachusetts Institute of Technology, Cambridge, MA 02139} 
\author{T.~Akdogan}
\affiliation{Laboratory for Nuclear Science and Bates Linear Accelerator
  Center, Massachusetts Institute of Technology, Cambridge, MA 02139} 
\author{H.~Arenh\"{o}vel}
\affiliation{Institut f\"{u}r Kernphysik, 
  Johannes Gutenberg-Universit\"{a}t Mainz, D-55099
  Mainz, Germany}
\author{R.~Alarcon}
\affiliation{Arizona State University, Tempe, AZ 85287}
\author{W.~Bertozzi}
\affiliation{Laboratory for Nuclear Science and Bates Linear Accelerator
  Center, Massachusetts Institute of Technology, Cambridge, MA 02139} 
\author{E.~Booth}
\affiliation{Boston University, Boston, MA 02215}
\author{T.~Botto}
\affiliation{Laboratory for Nuclear Science and Bates Linear Accelerator
  Center, Massachusetts Institute of Technology, Cambridge, MA 02139} 
\author{J.~Calarco}
\affiliation{University of New Hampshire, Durham, NH 03824}
\author{B.~Clasie}
\affiliation{Laboratory for Nuclear Science and Bates Linear Accelerator
  Center, Massachusetts Institute of Technology, Cambridge, MA 02139} 
\author{C.B.~Crawford}
\affiliation{Laboratory for Nuclear Science and Bates Linear Accelerator
  Center, Massachusetts Institute of Technology, Cambridge, MA 02139}  
\author{A.~DeGrush}
\affiliation{Laboratory for Nuclear Science and Bates Linear Accelerator
  Center, Massachusetts Institute of Technology, Cambridge, MA 02139} 
\author{T.W.~Donnelly}
\affiliation{Laboratory for Nuclear Science and Bates Linear Accelerator
  Center, Massachusetts Institute of Technology, Cambridge, MA 02139} 
\author{K.~Dow}
\affiliation{Laboratory for Nuclear Science and Bates Linear Accelerator
  Center, Massachusetts Institute of Technology, Cambridge, MA 02139} 
%\author{D.~Dutta}
%\affiliation{Triangle Universities Nuclear Laboratory and Duke University,
%  Durham, NC 27708-0305}
\author{M.~Farkhondeh}
\affiliation{Laboratory for Nuclear Science and Bates Linear Accelerator
  Center, Massachusetts Institute of Technology, Cambridge, MA 02139} 
\author{R.~Fatemi}
\affiliation{Laboratory for Nuclear Science and Bates Linear Accelerator
  Center, Massachusetts Institute of Technology, Cambridge, MA 02139} 
\author{O.~Filoti}
\affiliation{University of New Hampshire, Durham, NH 03824}
\author{W.~Franklin}
\affiliation{Laboratory for Nuclear Science and Bates Linear Accelerator
  Center, Massachusetts Institute of Technology, Cambridge, MA 02139} 
\author{H.~Gao}
\affiliation{Triangle Universities Nuclear Laboratory and Duke University,
  Durham, NC 27708-0305}
\author{S.~Gilad}
\affiliation{Laboratory for Nuclear Science and Bates Linear Accelerator
  Center, Massachusetts Institute of Technology, Cambridge, MA 02139} 
%\author{W.~Haeberli}
%\affiliation{University of Wisconsin, Madison, WI 53706}
\author{D.~Hasell}
\affiliation{Laboratory for Nuclear Science and Bates Linear Accelerator
  Center, Massachusetts Institute of Technology, Cambridge, MA 02139} 
%\author{W.~Hersman}
%\affiliation{University of New Hampshire, Durham, NH 03824}
%\author{M.~Holtrop}
%\affiliation{University of New Hampshire, Durham, NH 03824}
\author{P.~Karpius}
\affiliation{University of New Hampshire, Durham, NH 03824}
\author{H.~Kolster}
\affiliation{Laboratory for Nuclear Science and Bates Linear Accelerator
  Center, Massachusetts Institute of Technology, Cambridge, MA 02139} 
\author{T.~Lee}
\affiliation{University of New Hampshire, Durham, NH 03824}
\author{A.~Maschinot}
\affiliation{Laboratory for Nuclear Science and Bates Linear Accelerator
  Center, Massachusetts Institute of Technology, Cambridge, MA 02139} 
\author{J.~Matthews}
\affiliation{Laboratory for Nuclear Science and Bates Linear Accelerator
  Center, Massachusetts Institute of Technology, Cambridge, MA 02139} 
%twelve lines until here
\author{K.~McIlhany}
\affiliation{United States Naval Academy, Annapolis, MD 21402}
\author{N.~Meitanis}
\affiliation{Laboratory for Nuclear Science and Bates Linear Accelerator
  Center, Massachusetts Institute of Technology, Cambridge, MA 02139} 
\author{R.G.~Milner}
\affiliation{Laboratory for Nuclear Science and Bates Linear Accelerator
  Center, Massachusetts Institute of Technology, Cambridge, MA 02139} 
\author{J.~Rapaport}
\affiliation{Ohio University, Athens, OH 45701}
\author{R.P.~Redwine}
\affiliation{Laboratory for Nuclear Science and Bates Linear Accelerator
  Center, Massachusetts Institute of Technology, Cambridge, MA 02139} 
\author{J.~Seely}
\affiliation{Laboratory for Nuclear Science and Bates Linear Accelerator
  Center, Massachusetts Institute of Technology, Cambridge, MA 02139} 
\author{A.~Shinozaki}
\affiliation{Laboratory for Nuclear Science and Bates Linear Accelerator
  Center, Massachusetts Institute of Technology, Cambridge, MA 02139} 
\author{S.~\v{S}irca}
\affiliation{Laboratory for Nuclear Science and Bates Linear Accelerator
  Center, Massachusetts Institute of Technology, Cambridge, MA 02139} 
\author{A.~Sindile}
\affiliation{University of New Hampshire, Durham, NH 03824}
\author{E.~Six}
\affiliation{Arizona State University, Tempe, AZ 85287}
\author{T.~Smith}
\affiliation{Dartmouth College, Hanover, NH 03755}
\author{M.~Steadman}
\affiliation{Laboratory for Nuclear Science and Bates Linear Accelerator
  Center, Massachusetts Institute of Technology, Cambridge, MA 02139} 
\author{B.~Tonguc}
\affiliation{Arizona State University, Tempe, AZ 85287}
\author{C.~Tschalaer}
\affiliation{Laboratory for Nuclear Science and Bates Linear Accelerator
  Center, Massachusetts Institute of Technology, Cambridge, MA 02139} 
\author{E.~Tsentalovich}
\affiliation{Laboratory for Nuclear Science and Bates Linear Accelerator
 Center, Massachusetts Institute of Technology, Cambridge, MA 02139} 
\author{W.~Turchinetz}
\affiliation{Laboratory for Nuclear Science and Bates Linear Accelerator
  Center, Massachusetts Institute of Technology, Cambridge, MA 02139} 
%\author{J.F.J.~van~den~Brand}
%\affiliation{Vrije Universiteit and NIKHEF, Amsterdam, The Netherlands}
%\author{J.~van~der~Laan}
%\affiliation{Laboratory for Nuclear Science and Bates Linear Accelerator
%  Center, Massachusetts Institute of Technology, Cambridge, MA 02139} 
%\author{F.~Wang}
%\affiliation{Laboratory for Nuclear Science and Bates Linear Accelerator
%  Center, Massachusetts Institute of Technology, Cambridge, MA 02139} 
%\author{T.~Wise}
%\affiliation{University of Wisconsin, Madison, WI 53706}
\author{Y.~Xiao}
\affiliation{Laboratory for Nuclear Science and Bates Linear Accelerator
  Center, Massachusetts Institute of Technology, Cambridge, MA 02139} 
\author{W.~Xu}
\affiliation{Triangle Universities Nuclear Laboratory and Duke University,
  Durham, NC 27708-0305}
\author{C.~Zhang}
\affiliation{Laboratory for Nuclear Science and Bates Linear Accelerator
  Center, Massachusetts Institute of Technology, Cambridge, MA 02139} 
\author{Z.~Zhou}
\affiliation{Laboratory for Nuclear Science and Bates Linear Accelerator
  Center, Massachusetts Institute of Technology, Cambridge, MA 02139} 
\author{T.~Zwart}
\affiliation{Laboratory for Nuclear Science and Bates Linear Accelerator
  Center, Massachusetts Institute of Technology, Cambridge, MA 02139} 

\collaboration{The BLAST Collaboration}
\noaffiliation

\date{\today}

\begin{abstract}
We report new measurements of the neutron charge form factor
at low momentum transfer using quasielastic electrodisintegration of the
deuteron.  Longitudinally polarized electrons at an energy of 850 MeV
%stored in the South Hall Ring (SHR) of the MIT-Bates Linear Accelerator Center
were scattered from an isotopically pure, highly polarized
deuterium gas target.
%internal to the storage ring. 
The scattered electrons
and coincident neutrons were measured by the Bates Large Acceptance
Spectrometer Toroid (BLAST) detector. The neutron form factor ratio
%$\mu_n G^{n}_{E}/G^{n}_{M}$
$G^{n}_{E}/G^{n}_{M}$ 
was extracted from the 
%spin-perpendicular 
beam-target vector asymmetry $A_{ed}^{V}$ at four-momentum transfers
$Q^{2}=0.14$, 0.20, 0.29 and 0.42 (GeV/c)$^{2}$. 
  
\end{abstract}

\pacs{13.40.-f, 13.40.Gp, 13.88.+e, 14.20.Dh, 25.30.Bf}
%Valid PACS appear here}
\keywords{Neutron, form factor, polarization, internal target, elastic,
  deuterium} 

\maketitle
%\section{Introduction}
The neutron is composed of charged constituents, whose net 
distribution
is described by the charge (or electric) form factor $G^n_E$. 
Differences in the up and down quark distributions produce
a nonuniform distribution of the net %neutron 
charge~\cite{thomasweise}. 
The neutron electric form factor $G^n_E$ exhibits a maximum in the region of 
$Q^2 \approx 0.1-0.5$ (GeV/c)$^2$; when Fourier-transformed this corresponds
to a positively charged core and a concentration of 
negative charge at intermediate to large distances of $\approx 1$~fm,
commonly associated with a meson cloud surrounding the nucleon. 
%Since the charges of the valence quarks in the neutron sum to zero, it is
%expected that the charge form factor at low $Q^2$ $\approx$ 0.1 (GeV/c)$^2$
%(i.e. at long distances $\approx$ 2 Fermi) should be substantial in magnitude
%and should be sensitive to the tail of the neutron charge
%distribution~\cite{tail}. 
Models which emphasize the role of the meson cloud
have been successful in explaining important aspects of nucleon
structure~\cite{meson, pasquini, lomon, belushkin, miller02}. 
%% does the above paragraph make any sense??
Precise knowledge of $G^n_E$ is essential for the description of
electromagnetic structure of nuclei, and for the interpretation of parity
violating electron scattering experiments to determine the strangeness content
of the nucleon. 
Further, it is anticipated that exact {\it ab initio} QCD calculations of 
$G^n_E$ using lattice techniques will eventually be
possible~\cite{alexandrou}.  

In the absence of a free neutron target,
%feasible for scattering experiments
%and because of the small value of $G^n_E$ relative to the neutron magnetic
%form factor $G^n_M$, 
determinations of $G^n_E$ at finite $Q^2$ are typically
carried out using quasielastic electron scattering from
%the neutron bound in
deuterium or $^3$He targets.  
Despite the small value of $G^n_E$ compared to the neutron magnetic form
factor $G_M^n$, it can be obtained with high precision from
double-polarization observables based on the interference of $G_E^n$ with 
$G^n_M$.  
With the availability of high-duty-factor polarized electron beams
over the last decade, experiments have employed recoil
polarimeters, %~\cite{recoil_pol}, 
and targets of polarized $^2$H %~\cite{pol_deut}
and $^3$He %~\cite{pol_he3} 
to perform precision measurements of $G^n_E$
using polarization techniques with inherently small systematic
uncertainties~\cite{GEn_data}. 
The slope of $G^n_E(Q^2)$ at $Q^2$ = 0, which defines
the square of the neutron charge radius, is determined precisely by the 
scattering of thermal neutrons from atomic electrons~\cite{gen_slope}.  

This paper reports on new measurements of 
%$\mu_n G^{n}_{E}/G^{n}_{M}$ 
$G^{n}_{E}/G^{n}_{M}$ 
at low
$Q^2$ in the vicinity of the maximum of $G^n_E$, using a longitudinally
polarized electron beam incident on a vector-polarized $^{2}$H target internal
to the South Hall Ring at the MIT-Bates Linear Accelerator Center.  The BLAST 
detector was used to detect quasielastically scattered electrons in
coincidence with recoil neutrons over a range of $Q^{2}$ between 0.10 and 0.55
(GeV/c)$^{2}$. 

The differential cross section for the 
%$^{2}{\rm H}({\rm e},{\rm e}'{\rm n}){\rm p}$ 
\Deen reaction
with polarized beam and target can be
written~\cite{cross_section,arenhoevel05,arenhoevel88} %\pagebreak 
%\begin{eqnarray}
% d^{3}\sigma/(d\Omega_{e}d\Omega_{pq}d\omega) & = & 
%  \sigma_{unp} (1 + \sqrt{\frac{3}{2}}P_{z}A^{V}_{d} +
%  \sqrt{\frac{1}{2}}P_{zz}A^{T}_{d} + \\
%  & &  P_eA_{e} + \sqrt{\frac{3}{2}}P_eP_{z}A^{V}_{ed} +
%  \sqrt{\frac{1}{2}}P_eP_{zz}A^{T}_{ed}),
%\end{eqnarray}
\begin{equation}
{ d^{3}\sigma/(d\Omega_{e}d\Omega_{pq}d\omega}) \; = \;
%{ \displaystyle \frac{d^{3}\sigma}{d\Omega_{e}d\Omega_{pq}d\omega}} \; = \;
  \sigma_{unp} (1+\Sigma + P_e \Delta) %, %{\rm, \quad with}
\end{equation}
with
\begin{equation}
  \begin{array}{rcl}
\Sigma & = & { \displaystyle \sqrt{\frac{3}{2}}P_{z}A^{V}_{d}+
  \sqrt{\frac{1}{2}}P_{zz}A^{T}_{d}} \\  
  \Delta & = & { \displaystyle A_{e}+\sqrt{\frac{3}{2}}P_{z}A^{V}_{ed}+
  \sqrt{\frac{1}{2}}P_{zz}A^{T}_{ed}},
  \end{array}
  \label{qelXS}
\end{equation}
where $\sigma_{unp}$ is the unpolarized differential cross section, 
$P_{z} = n_{+}- n_{-}$ and $P_{zz} = n_{+} + n_{-} - 2n_{0}$ are the vector
and tensor polarizations of the deuteron target defined by the relative
populations $n_m$ of the three deuteron magnetic substates with
respect to the deuteron orientation axis, $m=+1,0,-1$,
respectively, and $P_e$ is the longitudinal polarization of the electron
beam. 
%The explicit coefficients $\sqrt{3/2}$ and $\sqrt{1/2}$ in
%Eq.~(\ref{qelXS}) arise from the density matrix description of the
%deuteron in the formalism of~\cite{arenhoevel05,arenhoevel88}. 

With BLAST, all of the polarization observables $A_i$ 
in Eq.~(\ref{qelXS}) have been
measured for the first time with precision in a single
experiment. 
%This fact allows to precisely determine the neutron form factor
%ratio and to validate the corrections due to FSI that are sizable at low
%$Q^2$. 
The beam-target vector polarization observable $A_{ed}^{V}$ 
is particularly sensitive to the neutron form factor ratio
$G^{n}_{E}/G^{n}_{M}$~\cite{arenhoevel88}.  
In the Plane Wave Born Approximation (PWBA) and with the deuteron in a pure 
S-state, the asymmetry $A_{ed}^{V}$ can be written analogously to elastic 
scattering from the free neutron as
\begin{equation}
  \begin{array}{ccc}
    { \displaystyle A_{ed}^{V}} & = & { \displaystyle 
      \frac{{a\, G^{n}_{M}}^{2} \cos\theta^{*} +
      b\, G^{n}_{E} G^{n}_{M}\sin\theta^{*}\cos\phi^{*}}{c\, {G^{n}_{E}}^{2} +
      {G^{n}_{M}}^{2}}} \\[14pt]  
   & \approx & { \displaystyle a \cos\theta^{*} +
  b\, \frac{G^{n}_{E}}{G^{n}_{M}} \sin\theta^{*} \cos\phi^{*}},
  \end{array}
  \label{asymm}
\end{equation}
where $\theta^{*}$ and $\phi^{*}$ are the target spin orientation angles
with respect to the momentum transfer vector and $a$, $b$, and $c$ are known
kinematic factors. This 
%polarization observable exhibits the maximum
asymmetry has the largest
sensitivity to $G_E^n$ when the momentum transfer
vector is perpendicular to the target polarization,
{\it i.e.} $\theta^* = 90^\circ$.   

\begin{figure}[ht]
  \centering
  \vspace*{-4ex}
  %\fbox{\rule[0cm]{0cm}{4cm}\rule[0cm]{.46\textwidth}{0cm}}
  %\fbox{\rule[0cm]{0cm}{4cm}\rule[0cm]{.46\textwidth}{0cm}}
  %\includegraphics[width=0.49\textwidth]{figures/Asym_32}
  \includegraphics[width=0.49\textwidth]{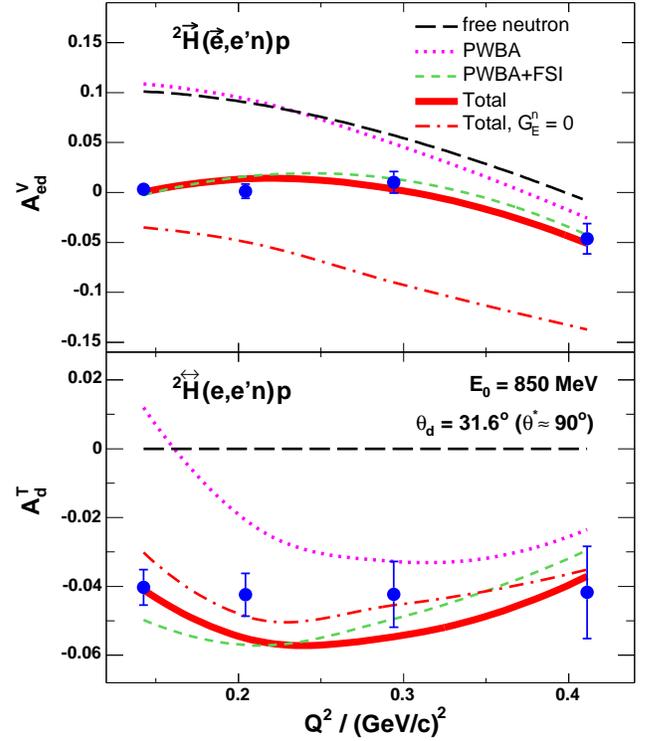}
  \caption[Sensitivity to the model]{Measured (solid blue points) and
  calculated beam-target vector polarization observable $A^{V}_{ed}$ (upper
  panel) and %\overleftrightarrow
  tensor asymmetry $A_d^T$ (lower panel) for the 
  %$^{2}\vec{\rm H}(\vec{\rm e},{\rm e}'{\rm n}){\rm p}$ and 
  %$^2\!\!\!\stackrel{\leftrightarrow}{\rm H}\!\!\!({\rm e},
  %{\rm e}'{\rm n}){\rm p}$ 
  %
  %$^{2}{\rm H}({\rm e},{\rm e}'{\rm n}){\rm p}$ 
  \Deenp$ $ 
  reaction at 850 MeV, a target
  orientation of $\theta_d = 31.6^\circ$ into the left sector of BLAST, 
  %where also the scattered electrons are detected. 
  and with neutrons detected in the right sector.
  The colored curves are Monte
  Carlo simulations based on the deuteron electrodisintegration model of
  Ref.~\cite{arenhoevel05} 
  %for successive inclusion of different reaction
  %mechanisms 
  (dotted magenta = PWBA, short-dashed green = PWBA+FSI,
  solid red = PWBA+FSI+MEC+IC+RC) using standard parameterizations for
  the nucleon form factors (see text). 
  In addition, the corresponding curves for $G_E^n \equiv 0$ 
  (dash-dotted red)
  and for elastic scattering from the free neutron (dashed black line) are
  shown. 
  %The agreement of the measured tensor asymmetry $A_d^T$ with the full model
  %supports the calculation of FSI for an accurate extraction of
  %$G_E^n$ from $A_{ed}^V$.
}  
  \label{model_sensitivity}
\end{figure}

However, there are sizable corrections to the asymmetry in Eq.~(\ref{asymm}),
mainly at low $Q^2$ where they are dominated by final state interactions
(FSI). The relative contributions of meson exchange currents (MEC), isobar
configurations (IC) and relativistic corrections (RC) become more significant
as the momentum transfer increases (see Fig.~\ref{model_sensitivity}). 
Extracting $G^{n}_{E}$ must be done by comparison with
theoretical asymmetries that include these effects. 

The effects of FSI can be monitored with the other polarization observables in
Eq.~(\ref{qelXS}). The asymmetries $A_e$, $A_d^V$, and $A_{ed}^T$ all vanish
in the Born approximation due to parity and time reversal conservation and
remain very small (below 1\%) even in the presence of FSI. This permits these
observables to be used to identify any false asymmetries in the experiment. 
FSI gives a sizable contribution to the target tensor asymmetry
$A_d^T$, 
%which would otherwise be expected to become zero in the quasifree 
%limit. 
which is insensitive to $G_E^n$ and otherwise close to zero in the quasifree
limit. 
Figure~\ref{model_sensitivity} displays a Monte Carlo simulation of the
reaction mechanism effects on the asymmetries $A_{ed}^V$ (upper panel) and
$A_d^T$ (lower panel) as a function of $Q^2$ along with the measured values.  
The calculations use the standard dipole form factor $G_D = (1+Q^2/0.71)^{-2}$
for $G_E^p$, $G_M^p/\mu_p$, and $G_M^n/\mu_n$, and %the Galster
%parameterization 
$1.91\tau/(1+5.6\tau) \,G_D$ for $G_E^n$~\cite{galster},
where $\mu_p = 2.79$, $\mu_n = -1.91$, and $\tau=Q^2/(4 m_n^2)$. 
%Comparing the calculations with the  
%measured tensor asymmetry $A_d^T$, the value of $\chi^2$/d.o.f. 
%improves from $117.66/4$ % 117.66/4 = 29.42
%in PWIA to $5.96/4$ % 5.96/4 = 1.49
%under inclusion of FSI, MEC, IC and RC. 
The good agreement of the measured tensor asymmetry $A_d^T$ with the full 
model supports the calculations of FSI for a reliable extraction of
$G_E^n$ from the beam-target vector asymmetry $A_{ed}^V$ at the percent
level. 

On the other hand, the corrections at low $Q^2$ to $A^{V}_{ed}$ measured in
the \vecDveceepn %$^2\vec{\rm H}$($\vec{\rm e}$,e'p)n 
%$^2\vec{\rm H}$($\vec{\rm e}$,e$^\prime$p)n 
reaction in quasifree kinematics are 
negligible~\cite{arenhoevel88}, which allows for a precise determination of
the product of beam and target polarizations $P_e P_{z}$ along with the proton 
form factor ratio $G^{p}_{E}/G^{p}_{M}$ in this reaction channel~\cite{deep}.

%\section{Experimental Details}
The BLAST experiment was designed 
%for the measurement of 
to carry out spin-dependent
electron scattering from hydrogen~\cite{crawford} and light nuclei.
% to an
%extensive set of fully reconstructed final states at an energy of 850
%MeV. 
%It utilizes the longitudinally polarized electron beam
%stored in the Bates South Hall Ring (SHR) and a polarized internal gas
%target. 
Details on the experimental setup can be found in~\cite{blastpaper}.
%, however the key features are briefly summarized in the following.
The internal target consisted of an atomic beam source (ABS) combined with an
open-ended storage cell through which the stored electron beam passed
continuously~\cite{abspaper}. 
The ABS produced polarized monoatomic deuterium
gas in the storage cell with nuclear vector ($V+$: $m$=1; $V-$: $m$=$-1$)
and tensor ($T-$: $m$=0) polarization states. In addition, the helicity $h$ of
the electron beam was flipped every injection cycle.
Linear combinations of the six charge-normalized yields $Y_{hm}$ 
define all five polarization observables in Eq.~(\ref{qelXS}). The
experimental value of the beam-vector polarization observable $A^{V}_{ed}$ can
be written as  
\begin{equation}
   A_{ed}^{V} =  
   \sqrt{\frac{3}{2}}\frac{1}{P_e P_{z}}
    \frac{Y_{++}+Y_{--}
      -Y_{+-}-Y_{-+}}{Y_{tot}},
\label{asymmexp}
\end{equation}
where $Y_{tot}$ % = \sum_{hm} Y_{hm}$ 
is the total yield obtained by summing up all six combinations $hm$.
A modest magnetic holding field was applied to define
the polarization angle $\theta_d$ within the horizontal plane
and to minimize the depolarization of target atoms. 
%The non-uniformity of $\theta_d$ along the target cell was
%carefully mapped with movable 
%Hall probes and a compass device. 
The variation of $\theta_d$ was carefully mapped over the extent of the target 
cell. The average value of $\theta_d$ was determined along with the tensor
polarization $P_{zz}$
%The absolute calibration of the average
%orientation angle $\theta_d$ along with a determination of the tensor
%polarization $P_{zz}$ was done 
by comparing the simultaneously
measured tensor asymmetries in elastic scattering from tensor-polarized
deuterium~\cite{zhang} 
with those expected at low $Q^2$ based on a parameterization of
previous data~\cite{abbott}.

The BLAST detector is a toroidal spectrometer (8 sectors) with the horizontal 
sectors instrumented with wire chambers, aerogel \v{C}erenkov counters, thin
plastic timing scintillators, and thick plastic scintillator walls for neutron
detection.
With the target polarization vector pointing into the left sector,
the neutron detection efficiency was augmented  
%five of the six walls with a resulting detector thickness of 30 cm were
%installed 
in the right sector covering the kinematic region most sensitive to
the neutron form factor ratio, as indicated by the $\sin\theta^{*}$ term in 
Eq.~(\ref{asymm}). The detection of neutrons in the left sector was
primarily used to independently verify the
determination of $P_e P_{z}$ from the \vecDveceepn
%$^2\vec{\rm H}$($\vec{\rm e}$,e'p)n
reaction.   
The selection of \een %(e,e'n) 
events is very clean; the number of proton tracks misidentified as neutrons is
negligible, due to the highly efficient charged particle veto provided by the
thin scintillator bars and the large-volume drift chambers in front of the
neutron detectors. 
A set of cuts applied on the time correlation between the 
charged and the neutral track, and on kinematic constraints for the
electrodisintegration process, was employed to identify the quasielastic
\een %(e,e'n) 
events.
The background from scattering off the aluminum target cell walls, measured
with a hydrogen (empty) target, is less than 4\% (3\%) 
of the normalized yield obtained with deuterium.

The corrected asymmetries were compared to Monte Carlo
simulations based on the deuteron electrodisintegration
model~\cite{arenhoevel05}, for which events were generated according to the
unpolarized cross section and weighted event-by-event with the
spin-dependent terms in Eq.~(\ref{qelXS}).
The acceptance-averaged asymmetry $A_{ed}^{V}$ was simulated for different
values of 
%$\mu_n G^{n}_{E}/G^{n}_{M}$ 
$G^{n}_{E}/G^{n}_{M}$ and compared to the experimental values. 
In order to extract the best value of the form factor ratio for each $Q^2$
bin,  
a $\chi^2$ minimization was performed independently with respect to the
missing momentum of the reaction and the angle of the  
neutron in the hadronic center-of-mass system. Both extractions produced
consistent results. 

%\section{Results}
The data reported here were acquired in two separate runs in 
2004 and 2005, corresponding to a target polarization angle of
%$31.64^\circ \pm 0.35^\circ ({\rm stat}) \pm 0.24^\circ ({\rm sys})$ 
%and $46.32^\circ \pm 0.42^\circ ({\rm stat}) \pm 0.16^\circ ({\rm sys})$, 
$31.64^\circ \pm 0.43^\circ$ and $46.32^\circ \pm 0.45^\circ$, 
respectively.
With a total accumulated beam charge of 451 kC (503 kC) in the first (second)
data set, final samples of 
%206,570 (148,006) 
%176,252 (156,190) 
268,914 (205,252) %before pm<0.2 and th>140 cut, both sectors
% 2004:
coincident electron-neutron
events were collected. The average product of beam and target polarization
determined from the \vecDveceep
%$^2 \vec{\rm H}$($\vec{\rm e}$,e'p) 
reaction was 
%$P_e P_{z} = 0.541 \pm 0.004 ({\rm stat}) \pm 0.0?? ({\rm sys})$ in the first
%and $0.468 \pm 0.005 ({\rm stat}) \pm 0.0?? ({\rm sys})$ in the second
$P_e P_{z} = 0.5796 \pm 0.0034 ({\rm stat}) \pm 0.0034 ({\rm sys})$ in the
first and $0.5149 \pm 0.0043 ({\rm stat}) \pm 0.0054 ({\rm sys})$ in the second
data set~\cite{deep}. 
In comparison, the polarization product determined from \vecDveceenp 
%$^2\vec{\rm H}$($\vec{\rm e}$,e'n)p 
with neutrons detected in the left sector
of BLAST corresponding to $\theta^* \approx 0^\circ$, was found
to be 
%$0.547 \pm 0.012$(stat) 
$0.587 \pm 0.019$(stat) 
%for the first
and 
%$0.481 \pm 0.019$(stat) 
$0.481 \pm 0.026$(stat) 
%for the second data set, 
consistent with the above \eep %(e,e'p) 
results.
%but less precise.  
The two data sets were treated as separate experiments producing two
consistent results for the form factor ratio, which were combined for a final
result. The data were divided into four $Q^2$ bins
%selected for  the determination of 
%$\mu_n G^{n}_{E}/G^{n}_{M}$ 
to determine
$G^{n}_{E}/G^{n}_{M}$ 
with a comparable statistical
significance
(see Table~\ref{tab:results}).
%(5\%).
%The uncertainty of the beam-target polarization product $P_e P_{z}$
%($xxx$\%) partially compensated by the uncertainty of the polarization angle. 
%$\mu_n G^{n}_{E}/G^{n}_{M}$
%The results are compiled in Table~\ref{tab:results}.
%\begin{table}[ht]
%  \centering
%  \begin{tabular}{c@{\quad}c@{$\pm$}c@{$\pm$}c@{\quad}c@{$\pm$}c@{$\pm$}c}
%    \hline\hline
%    & 
%    \multicolumn{3}{c@{\quad}}{2004} & 
%    \multicolumn{3}{c@{\quad}}{2005} \\ 
%    \hline
%    $\theta_d$ & %\pm{\rm stat}\pm{\rm sys}$ & 
%    31.64$^\circ$  & 0.35$^\circ$ & 0.24$^\circ$ & 46.32$^\circ$ & 0.24$^\circ$ & 0.16$^\circ$\\
%    $P_eP_z$ (e,e'p) & 0.5796 & 0.0034 & 0.0034 & 0.5149 & 0.0043 & 0.0054\\
%    $P_eP_z$ (e,e'n) & 0.587 & 0.019 & & 0.481 & 0.026\\
%    $Q_{total}$/kC\\
%    $N$ (e,e'n)\\
%    \hline
%  \end{tabular}
%\caption{Parameters of the BLAST experiment.}
%\label{tab:parameters}
%\end{table}
\begin{table}[b]
  \centering
  \begin{tabular}{r@{$-$}lc@{\quad}c@{$\;\pm\;$}c@{$\;\pm\;$}c}
%    \hline\hline
    \hline
    \multicolumn{2}{c@{\quad}}{$Q^2$/(GeV/c)$^2$} & 
    \multicolumn{1}{c@{\quad}}{$\langle Q^2 \rangle$/(GeV/c)$^2$} & 
    \multicolumn{3}{c@{\quad}}{$\mu_n G_E^n/G_M^n$} \\ 
    \hline
    0.10 & 0.17  & 0.142  & 0.0505 & 0.0072 & 0.0031\\
    0.17 & 0.25  & 0.203  & 0.0695 & 0.0084 & 0.0039\\
    0.25 & 0.35  & 0.291  & 0.1022 & 0.0127 & 0.0046\\
    0.35 & 0.55  & 0.415  & 0.1171 & 0.0182 & 0.0052\\
    \hline
  \end{tabular}
\caption{Results for the extracted neutron form factor ratio 
  $\mu_n G_E^n/G_M^n$ ($\mu_n = G_M^n(0) = -1.91$)
  with statistical and systematic errors, respectively.}
% The individual
%  systematic uncertainties were added in quadrature.}
\label{tab:results}
\end{table}

The systematic error of $G^{n}_{E}/G^{n}_{M}$
is dominated by the uncertainty of the target spin angle $\theta_d$.
Other systematic uncertainties include that of the beam-target polarization
product $P_e P_{z}$, % ($xxx$\%), 
the accuracy of kinematic reconstruction, % ($<$$xxx$\%),
as well as the dependency on software cuts. % ($xxx$\%). 
The systematic uncertainties were evaluated individually for each $Q^2$ 
bin and data set by combining the errors from each source, taking
covariances into account;  
%the systematic error of $P_eP_z$ is mostly due to the uncertainty in
%$theta_d$; 
the correlated and uncorrelated error categories of the two 
measurements were then combined for a resulting systematic error of each bin. 
False asymmetries were studied with the observables $A^{V}_{d}$ and
$A^{T}_{ed}$
%at the percent level
and found to be consistent with
zero. Radiative corrections to the asymmetries calculated in a PWBA formalism 
using the code MASCARAD~\cite{afanasev} are $<$$1$\% and therefore also
neglected.  
%All systematic
%uncertainties were added in quadrature, except for the uncertainties of
%$\theta_d$ and $P_e P_z$ which are partially correlated, thereby compensating
%a part of the systematic error. 
The uncertainties of the reaction mechanism and FSI corrections, which are
small compared to the experimental errors, are not included
in the systematic error.

\begin{figure}[t]
  \centering
 \vspace{-3ex}
 \includegraphics[width=.54\textwidth]{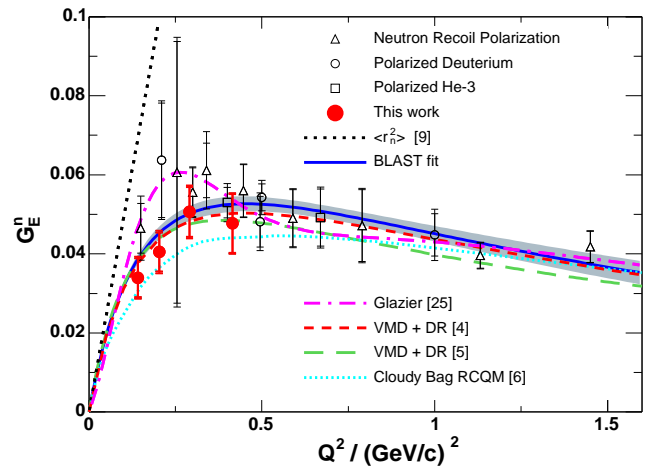}
 \caption{World data on $G_E^n$ from double-polarization
   experiments~\cite{GEn_data}. 
   %Only experiments with errors smaller than 30\% are shown.
   The data correspond to neutron recoil polarization experiments with
   unpolarized $^2$H %deuterium 
   target %~\cite{recoil_pol} 
   (open triangles) and experiments with polarized 
   $^2$H %deuterium 
   %target 
   (open circles; % = Ref.~\cite{pol_deut}, 
   solid red dots = this work)
   and %polarized 
   $^3$He %helium-3
   targets %~\cite{pol_he3} 
   (open squares).
   %, and from the analysis of the deuteron quadrupole form factor
   %GQ~\cite{schiavilla_and_sick} (open diamonds). 
   The data are shown with statistical (small error bars) and with statistical
   and systematic errors added quadratically (large error bars). 
   The ``BLAST fit'' (blue solid line) is a parameterization of the data 
   based on the sum of two dipoles
   %$\sum_{1,2}a_{1,2}/(1+Q^2/b_{1,2})^2$ 
   shown with a one-sigma error band. The recent
   parameterization~\cite{glazier} %``Mainz fit'' 
   (magenta dash-dotted line) is based on the form introduced
   in~\cite{friedrich_walcher}. 
   Also shown are recent results based on vector meson dominance and
   dispersion relations (red short-dashed~\cite{lomon} and 
   green long-dashed lines~\cite{belushkin}),
   and of a light-front cloudy bag model with relativistic constituent
   quarks~\cite{miller02} (cyan dotted line).
}
\label{gen}
\end{figure}

%\section{Discussion}
The world's data on $G^{n}_{E}$ from double-polarization
experiments~\cite{GEn_data}
%~\cite{recoil_pol,pol_deut,pol_he3} 
are displayed in Fig.~\ref{gen}  
along with the results of this work. 
All of the polarization data were experimentally determined as electric to
magnetic form factor ratios. 
%Typically, the standard dipole parameterization
%is chosen for the magnetic form factor in order to extract
%$G^{n}_{E}$. Instead, we used a more realistic
We used 
parameterization~\cite{friedrich_walcher} for $G_M^n$, which is in good
agreement with recent measurements~\cite{GMn_data}, %~\cite{gmn_experiments}, 
%in the low $Q^2$ region 
to determine $G_E^n$ from BLAST and to adjust the previously published
values. The data from a variety of experiments are consistent and
remove the large model uncertainty of previous $G_E^n$ extractions from
elastic electron-deuteron scattering~\cite{platchkov}. The new distribution is
also in agreement with $G_E^n$ extracted from the deuteron quadrupole form 
factor~\cite{schiavilla_and_sick}. 

The measured distribution 
of $G_E^n$ can be parameterized as a function of
$Q^2$ based on the sum of two dipoles,
%$\sum_{1,2} a_{1,2}/(1+Q^2/b_{1,2})^2$
$\sum_i a_{i}/(1+Q^2/b_{i})^2$ ($i\!\!=\!\!1,2$),
%which is 
shown as the BLAST fit in Fig.~\ref{gen} (blue line) with a one-sigma 
error band. With $G_E^n(0)=0$ and the slope at $Q^2=0$ constrained by $\left <
  r_n^2 \right > = (-0.1148 \pm 0.0035)$ fm$^2$~\cite{gen_slope},
one parameter is fixed, 
%resulting in $a_1 = 0.151 \pm 0.024$, $b_1 = 2.157
%\pm 0.555$ and ${\rm cov}(a_1,b_1) = -0.013$.
%
%resulting in $a_1 = -a_2 = 0.089 \pm 0.020$, $b_1 = 2.84 \pm 0.97$, 
%$b_2 = 0.276 \pm 0.085$ and ${\rm cov}(a_1,b_1) = -0.019$, 
%${\rm cov}(a_1,b_2) = 0.0016$, ${\rm cov}(b_1,b_2) = -0.07$.
%
%esulting in $a_1 = -a_2 = 0.092 \pm 0.024$, $b_1 = 2.786 \pm 1.029$, 
%$b_2 = 0.298 \pm 0.100$ and ${\rm cov}(a_1,b_1) = -0.024$, 
%${\rm cov}(a_1,b_2) = 0.0023$, ${\rm cov}(b_1,b_2) = -0.090$.
%
%resulting in $a_1 = -a_2 = 0.106 \pm 0.023$, $b_1 = 2.369 \pm 0.684$, 
%$b_2 = 0.364 \pm 0.054$ and ${\rm cov}(a_1,b_1) = -0.015$, 
%${\rm cov}(a_1,b_2) = 0.0012$, ${\rm cov}(b_1,b_2) = -0.035$.
%
%resulting in $a_1 = -a_2 = 0.094 \pm 0.017$, $b_1 = 2.777 \pm 0.798$, 
%$b_2 = 0.336 \pm 0.045$ and ${\rm cov}(a_1,b_1) = -0.013$, 
%${\rm cov}(a_1,b_2) = 0.0007$, ${\rm cov}(b_1,b_2) = -0.033$.
%
%resulting in $a_1 = -a_2 = 0.094 \pm 0.017$, $b_1 = 2.782 \pm 0.800$, 
%$b_2 = 0.336 \pm 0.045$ and ${\rm cov}(a_1,b_1) = -0.013$, 
%${\rm cov}(a_1,b_2) = 0.0007$, ${\rm cov}(b_1,b_2) = -0.033$.
%
%resulting in $a_1 = -a_2 = 0.097 \pm 0.018$, $b_1 = 2.667 \pm 0.773$,   
%$b_2 = 0.343 \pm 0.048$ and ${\rm cov}(a_1,b_1) = -0.014$,  
%${\rm cov}(a_1,b_2) = 0.0009$, ${\rm cov}(b_1,b_2) = -0.034$ with $Q^2$ in
%
resulting in $a_1 = -a_2 = 0.095 \pm 0.018$, $b_1 = 2.77 \pm 0.83$,   
$b_2 = 0.339 \pm 0.046$ and ${\rm cov}(a_1,b_1) = -0.014$,  
${\rm cov}(a_1,b_2) = 0.0008$, ${\rm cov}(b_1,b_2) = -0.036$ with $Q^2$ in
units of (GeV/c)$^2$. The parameterization~\cite{glazier} (magenta dash-dotted 
line) is based on the 
form introduced in~\cite{friedrich_walcher} with an additional bump structure 
around $0.2-0.4$ (GeV/c)$^2$. Also shown are recent results based on vector
meson dominance (VMD) and dispersion relations
(red short-dashed~\cite{lomon} and green long-dashed lines~\cite{belushkin}),
and the prediction of a light-front cloudy bag model with relativistic
constituent quarks~\cite{miller02} (cyan dotted line). 

%While there is no statistical significance to either prefer the BLAST fit
%(a smooth distribution) or a form with an additional bump structure,
%it is noticed that both the dispersion analysis and the meson-cloud model
%seem to systematically predict slightly lower values of $G_E^n$.
The new data from BLAST do not show a %pronounced 
bump structure at low $Q^2$ as previously
suggested~\cite{friedrich_walcher,glazier}.  
The BLAST data are in excellent agreement with VMD based
models~\cite{lomon,belushkin}
and also agree with the meson-cloud calculation~\cite{miller02}.  
The improved precision of the data at low $Q^2$ provides strong
constraints on the theoretical understanding of the nucleon's meson cloud.  

We thank the staff of the MIT-Bates Linear Accelerator Center for delivering
high quality electron beam and for their technical support, and A. Bernstein
for suggesting the form of the BLAST fit. 
This work has been supported in part by the US Department of Energy
and National Science Foundation.
%under Cooperative Agreement DE-FC02-94ER40818.
 
\end{document}